# Spin-dependent optical response of multiferroic EuO

Wen-Yi Tong,[1] Hang-Chen Ding,[1] Yong-Chao Gao,[1] Shi-jing Gong,[1] Xiangang Wan,[2] and Chun-Gang Duan[1,3,*]

[1]*Key Laboratory of Polar Materials and Devices, Ministry of Education, East China Normal University, Shanghai 200062, China*

[2]*National Laboratory of Solid State Microstructures and Department of Physics, Nanjing University, Nanjing 210093, China*

[3]*National Laboratory for Infrared Physics, Chinese Academy of Sciences, Shanghai 200083, China*

Using first-principles density functional calculations, electronic and optical properties of ferromagnetic semiconductor EuO are investigated. In particular, we have developed a way to obtain the spin-dependent optical response of the magnetic materials, which is helpful to verify the spin-dependent band structure of EuO. Significantly different optical responses from spin-up and spin-down channels are obtained in both linear and nonlinear cases, making it possible to distinguish contributions from different spin-channels in the optical absorption spectra if spin-flip process can be neglected. In addition, the red-shift of the absorption edge from paramagnetic to ferromagnetic ordering is explained by exchange interactions. Using such method, we have also compared the optical properties of multiferroic EuO which is induced by strong epitaxial strain. Our results show that from tensile to compressive strain, the blue-shift of the leading absorption peaks in the optical spectra, the red-shift of the optical band gap in spin-up state can be observed, consistent to the energy difference between spin-splitting orbits. The spin-dependent nonlinear optical properties reveal that in the infrared and visible light region, the contributions to second-harmonic generation (SHG) susceptibilities are mainly from spin-majority channels. In addition, the strain effect is also discussed. With the increase of epitaxial strain, the larger energy shift of the leading absorption peaks, and the more remarkable nonlinear optical response can be obtained.

## I. INTRODUCTION

For many years, the rare-earth compound Europium oxide (EuO) has attracted wide attention due to its remarkable semiconducting and ferromagnetic (FM) properties.[1,2] EuO has a rocksalt crystal structure[3] with FM ordering which is driven by the Heisenberg exchange coupling between the localized Eu-4$f$ moment of 6.9 $\mu_B$/Eu below its Curie temperature ($T_C$) of 69K.[4] The pronounced ferromagnetism induces outstanding colossal magnetoresistance effect,[5] metal-to-insulator transition (MIT)[6] and perfect spin polarization close to the conduction band (CB) edge.[7,8] The ability of epitaxially integration with Si,[9,10] GaAs[11] and GaN[9] as well as much enhanced Curie temperature via electrons injecting,[12-18] isostatic pressure,[19-22] or epitaxial stain,[23,24] makes EuO exceptional and of interest for semiconductor-based spintronics[25] such as spin-filter tunnel barriers.[26-29] In addition, due to large Faraday[30,31] and Kerr[32] effects derived from the 4$f$-5$d$ electronic transition and spin configuration of the divalent Eu ions in EuO, considerable attention has been focused on its promising applications in magneto-optic devices.[33,34]

Despite of these tremendous observations in the field of linear optics, nonlinear optical properties of EuO, such as sum and difference frequency generation remained unexplored until very recently. Compared with linear optical techniques which are used to probe the magnetization, nonlinear optics is practical as a tool for providing complementary information on crystallographic, electronic and magnetic properties and studying the coexistence and interactions of magnetic and electric order.[35,36] Matsubara et al.[37] observed magnetization-induced optical second-harmonic generation (SHG), the leading order of nonlinear optical susceptibility, in epitaxial films of the intrinsic centrosymmetric FM semiconductor EuO. In this case, spectrally and spatially resolved SHG, which is dominated by magnetic-dipole contributions in the expansion of the electromagnetic light field, disclose a submicron size of FM domains. Soon after that, electric-dipole induced third-harmonic generation (THG) spectroscopic studies, as well as the research of giant third-order magneto-order rotation in out-of-plane-magnetized EuO films, have been performed by Matsubara and his colleague.[38,39] The nonlinear optical research mentioned above is in the systems with symmetric center. However, in the majority of cases, SHG is limited to systems where the inversion symmetry is locally broken. To some extent, the centrosymmetric cubic structure restricts the research of nonlinear optics for EuO. Fortunately, with the development of strain engineering, Bousquet et al.[40] predicted highly epitaxial strained EuO to become ferroelectric (FE) and pointed out that its magnetic state remains FM through the FE region. The detailed mechanism of induced ferroelectricity in EuO is investigated by Kim.[41] Although the strain is too large to be experimentally applied in EuO at present, the finding offers possibilities for the study of EuO as an attractive high-temperature multiferroic material together with the research of nonlinear optical response in noncentrosymmetric EuO.

In this paper, we investigate the spin-dependent linear optical properties of unstrained EuO. In most cases, regular spin-independent optical measurements can be carried out to explore band structures without the complication of external fields, which is of great value because the application of fields may transform inherent characteristics of materials. In fact, there have been many successful precedents.[42-45] For a FM semiconductor like EuO, electric and magnetic properties are extremely sensitive to spin state configurations. Calculated spin-dependent spectrum, together with detailed energy difference analysis, are necessary and adopted for FM EuO to research electron transitions among orbits. When compressive and tensile strain is applied, transition from paraelectric (PE) to FE phase occurs, which further triggers the change in linear optical response and the study of nonlinear optical properties. Especially, the linear dielectric function and SHG susceptibilities can be spin distinguished.

## II. COMPUTATIONAL METHOD

The calculations are performed within density-functional theory (DFT) using the accurate full-potential projector augmented wave (PAW) method, as implemented in the Vienna *ab initio* Simulation Package (VASP).[46-48] The exchange-correlation potential is treated in Perdew-Burke-Ernzerhof (PBE) form of the generalized gradient approximation (GGA)[49] with a kinetic-energy cutoff of 750 eV. A 6×6×6 and 12×12×12 Γ centered $k$-point grid are adopted in the self-consistent and optical calculations. The Brillouin zone integrations are calculated using the tetrahedron method with Blöchl corrections.[50] In order to give a better description of the partially filled and strongly correlated localized Eu-$f$ orbit, the LSDA+$U$ method is adopted.[51] According to Paul's work,[52] here the on-site Coulomb interaction parameter $U$ and exchange interaction parameter $J$ are chosen to be 7.397 and 1.109 eV, respectively. The structures are relaxed until the Hellmann-Feynman forces on each atoms are less than 1 meV/Å. The polarizations of the FE phase are obtained using the Berry phase method.[53]

For the optical property calculation, we adopt our own code OPTICPACK. It has been successfully used to study the optical, especially the nonlinear optical properties of the borates,[54,55] ferroelectric polymer[56] and carbon nanotubes.[57] Here, to research the linear optical response and separate the contributions from different spin channels, we rewrite the some important formulas related to the optical calculation into spin-dependent ones as follows.

First, the complex dielectric function is written as:

$$\varepsilon = \varepsilon_1 + i\varepsilon_2 = 1 + 4\pi\chi, \quad (1)$$

here $\chi$ is the electric susceptibility. The electric contribution to $\chi$ (high frequency part) contains linear and nonlinear part. Considering up to the second order, we have

$$\chi(\omega) = \chi^{(1)}(\omega) + \chi^{(2)}(\omega). \quad (2)$$

The imaginary part of $\chi^{(1)}$ equals to $\varepsilon_2/4\pi$, and $\varepsilon_2$ is related to the optical conductivity $\sigma$ using the following relation:

$$\varepsilon_2^{\uparrow(\downarrow)}(\omega) = \frac{4\pi}{\omega}\sigma^{\uparrow(\downarrow)}(\omega). \quad (3)$$

Here the up and down arrows in the equation refer to the spin-up and spin-down channel, respectively. Then we have written the formula into spin-dependent form and the final physical quantities are the sums from both spin channels. We can further rewrite the interband optical conductivity tensor $\sigma$ from the non-spin-polarized form[58] to spin-polarized one as (in atomic units):

$$\sigma^{\uparrow(\downarrow)} = \frac{\pi}{\omega\Omega}\sum_{\vec{k}} W_{\vec{k}} \sum_{c,v} \left|\left\langle c^{\uparrow(\downarrow)}\left|\vec{e}\cdot\vec{p}\right|v^{\uparrow(\downarrow)}\right\rangle\right|^2 \delta(E_c^{\uparrow(\downarrow)} - E_v^{\uparrow(\downarrow)} - \omega), \quad (4)$$

where $\omega$ is the photon energy, $\Omega$ is the cell volume, $\vec{e}$ is the polarization direction of the photon, and $\vec{p}$ its electron momentum operator. The integral over the $k$ space has been replaced by a summation over special $k$ points with corresponding weighting factor $W_{\vec{k}}$. The second summation includes the valence band (VB) states ($v$) and CB states ($c$), based on the reasonable assumption that the VB is fully occupied, while the CB is empty. The real part of $\chi^{(1)}$ can be obtained spin-dependently using the Kramers-Kronig (KK) transformation.

For the nonlinear case, at this stage we only consider the SHG coefficient $\chi^{(2)}(2\omega, \omega, \omega)$. Similarly, we have extended the expression $\chi^{(2)}$ from the non-spin-polarized[56] to spin-dependent form. Especially, the spin-dependent expression for $\chi^{(2)}$ in the static limit can be written as:

$$\chi_{ijk}^{(2)\uparrow(\downarrow)}(0) =$$
$$\frac{1}{2}\int_{BZ}\frac{dk}{4\pi^3}P^{\uparrow(\downarrow)}(ijk)\left[\sum_{vv'c}\mathrm{Im}\left(p_{v'v}^{i\uparrow(\downarrow)}p_{vc}^{j\uparrow(\downarrow)}p_{cv'}^{k\uparrow(\downarrow)}\right)\left(\frac{1}{\omega_{cv}^3\omega_{cv'}^2}+\right.\right.$$
$$\left.\left.\frac{2}{\omega_{cv}^4\omega_{cv'}}\right)+\sum_{vcc'}\mathrm{Im}\left(p_{cv}^{i\uparrow(\downarrow)}p_{vc'}^{j\uparrow(\downarrow)}p_{c'c}^{k\uparrow(\downarrow)}\right)\left(\frac{1}{\omega_{cv}^3\omega_{c'v}^2}+\frac{2}{\omega_{cv}^4\omega_{c'v}}\right)\right], \quad (5)$$

in which $i$, $j$, $k$ are Cartesian components. $P^{\uparrow(\downarrow)}(ijk)$ denotes full permutation over $i$, $j$, $k$ and explicitly shows Kleinman symmetry of the SHG coefficients. $v$ ($v'$) and $c$ ($c'$) represent VB and CB, respectively. $p_{mn}$ is the electron momentum matrix element between $m$ and $n$ states, and the energy difference between the two states is denoted as $\hbar\omega_{mn}$, which is indeed also spin-dependent.

## III. RESULTS AND DISCUSSION

To understand the salient features of the absorption spectra, and further accurately distinguish spin-dependent contributions to the optical response, the energy band structure, as well as the density of states (DOS) are needed to be inspected. As an example, in Fig. 1, we show these information for the case of unstrained EuO at $a$ = 5.184 Å with cubic binary rocksalt structure ($Fm3m$) in the FM phase. The top edge of the VB is three-band degenerate and located in the zone center. It is derived primarily from Eu-4$f$ spin-majority orbits. Whereas, the bottom edge of the CB is located at X point, consisting mainly from Eu-5$d$ spin-majority orbits. The direct (optical) band gap is about 0.75 eV at X point. For the spin-minority channel, the energy gap is much larger (3.4 eV), also at X point.

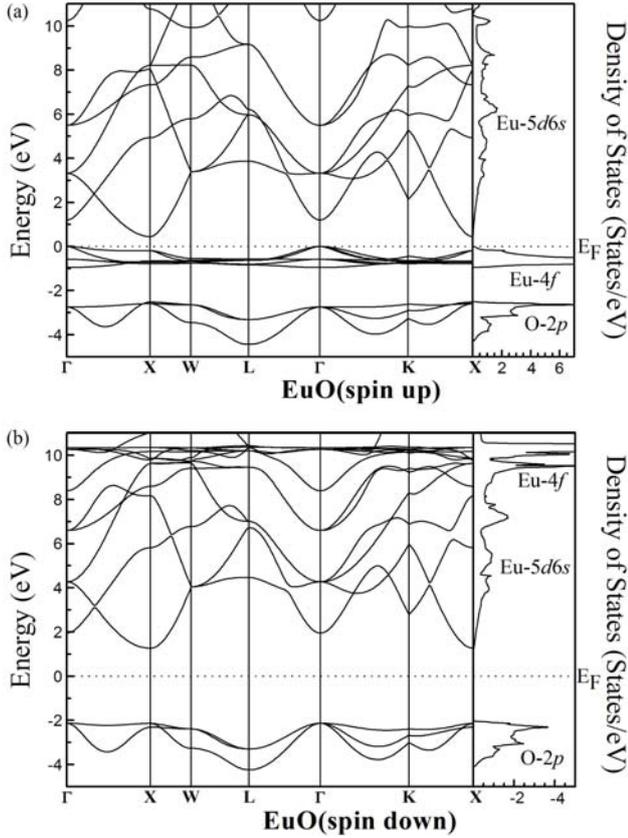

FIG. 1. (a) Spin-up and (b) spin-down band structure (left panel) and corresponding density of states (right panel) of unstrained EuO in the FM configuration with optimized bulk lattice parameters, $a$ = 5.184 Å. The zero of energy is at the Fermi level, $E_F$.

TABLE I. Calculated absorption peaks and the leading interband transitions of each peaks.

| Peaks | Spin | leading transition |
|---|---|---|
| A(2.68 eV) | ↑ | Eu-$f$ → Eu-5$d$6$s$ |
| B(4.12 eV) | ↑ | Eu-$f$ → Eu-5$d$6$s$ |
| C(4.54 eV) | ↑ | O-$p$ → Eu-5$d$6$s$ |
| D(5.46 eV) | ↑ | Eu-$f$ → Eu-5$d$6$s$ |
| E(6.98 eV) | ↑ | O-$p$ → Eu-5$d$6$s$ |
| F(8.80 eV) | ↑ | O-$p$ → Eu-5$d$6$s$ |
| G(7.70 eV) | ↓ | O-$p$ → Eu-5$d$6$s$ |
| H(9.48 eV) | ↓ | O-$p$ → Eu-5$d$6$s$ |
| I(10.64 eV) | ↓ | O-$p$ → Eu-5$d$6$s$ |

As the imaginary part of complex dielectric function $\varepsilon_2$ is closely related to the linear optical absorption, it has been calculated and shown in Fig. 2. Furthermore, the origin of the absorption peaks is analyzed spin-dependently as follows. We check that our results are robust within the representative choice of the on-site Coulomb interaction parameter $U$ and exchange interaction parameter $J$ used for Eu and the similar group-V element Gd, i.e. $U$ = 6.7 eV and $J$ = 0.7 eV,[59-61] $U$ = 7 eV and $J$ = 1.2 eV.[22] As the splitting of the occupied and unoccupied Eu-$f$ levels is given by the difference of the ($U_{eff} = U - J$), the optical responses are insensitive to the value of $U$ and $J$ when the difference $U_{eff}$ is at reasonable range.

For the spin-up case (up triangle line with red color), the weak peak A around 2.68 eV, the peak B around 4.12 eV and the peak D around 5.46 eV are all of majority Eu-$f$ to Eu-5$d$6$s$ transitions. The peak C around 4.54 eV, the dominant central peak E around 6.98 eV and the lower-energy spectral peak F around 8.80 eV are mainly attributed to transitions from O-$p$ to Eu-5$d$6$s$ majority states. For the spin-down case (down triangle line with blue color), all of the three peaks, peak G in 7.70 eV, peak H in 9.48 eV and peak I in 10.64 eV are due to significant contribution of minority O-$p$ to Eu-5$d$6$s$ transitions. The leading interband transitions of each absorption peaks are listed in Table I.

As is clear from Fig. 2, electrons with different spin have dramatically different dielectric response. This can be explained by their different band structures, especially the energy levels of Eu-4$f$ states, as shown in Fig. 1. Such information generally is hard to be obtained experimentally. With the help of our spin-dependent spectroscopy calculation, however, an ordinary optical measurement, can reveal relevant information about the spin-dependent band structure.

It is interesting to point out that detailed analysis reveals that there could exist tiny energy difference between the absorption peaks for total (solid square line with black color) and its corresponding spin states, when the optical absorption contains contributions from both spin channels, i.e., the photon energy is larger than the spin-down energy gap (here 3.4 eV). As a rough explanation of such energy shift, we take the derivative of the identity $\varepsilon_2^{tot} = \varepsilon_2^\uparrow + \varepsilon_2^\downarrow$, and obtain the equation $\frac{\partial \varepsilon_2^{tot}}{\partial E} = \frac{\partial \varepsilon_2^\uparrow}{\partial E} + \frac{\partial \varepsilon_2^\downarrow}{\partial E}$. The peaks in the total absorption curve occur only when the relationship $\frac{\partial \varepsilon_2^\uparrow}{\partial E} = -\frac{\partial \varepsilon_2^\downarrow}{\partial E}$ holds. When the photon energy is in the range of 3.4-7.5 eV and 8.7-9.4 eV, where $\varepsilon_2^\downarrow$ is in the uphill and therefore $\frac{\partial \varepsilon_2^\downarrow}{\partial E} > 0$, the peaks in the total absorption curve can only occur in the downhill of $\varepsilon_2^\uparrow$, where $\frac{\partial \varepsilon_2^\uparrow}{\partial E} < 0$. As a result, the peak positions B'(4.22 eV), C'(4.60 eV), D'(5.48 eV), E'(7.02 eV) and F'(8.84 eV) in the total curve are slightly larger than those corresponding peak positions, i.e. B(4.12 eV), C(4.54 eV), D(5.46 eV), E(6.98 eV) and F(8.80 eV), in the spin-up curve. It is noting that the greater the slope of $\varepsilon_2^\downarrow$ is, the larger energy difference between the absorption peak for total and spin-up case will be. Analogously, the negative $\frac{\partial \varepsilon_2^\uparrow}{\partial E}$ ( $\varepsilon_2^\uparrow$ is downhill) and positive $\frac{\partial \varepsilon_2^\downarrow}{\partial E}$ result in smaller energy of peaks G'(7.58eV), H'(9.42 eV), I'(10.58 eV) in total curve, compared with peaks G(7.70eV), H(9.48 eV) and I(10.64 eV) in the spin-down case.

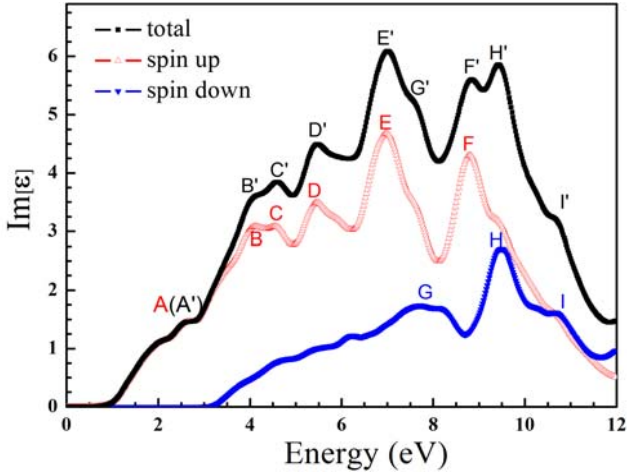

FIG. 2. (Color online) The spin-dependent imaginary part of complex dielectric function $\varepsilon$ for unstrained EuO. The solid square line with black color, up triangle line with red color and down triangle line with blue color are for total, spin-up and spin-down states, respectively.

All the above calculations and analyses are performed under FM ordering. Unfortunately, to our knowledge, though the spin-independent experimental spectra of EuO under paramagnetic (PM) ordering was researched over 40 years ago,[62] there is few related work on EuO in FM ordering. Then it is interesting to discuss the change of the linear optical response when the temperature reaches over $T_C$, i.e. when the system undergoes FM-PM transition?

As is known that the FM ordering of EuO is due to the exchange interaction of the Eu-4$f$ electrons. The local exchange interaction, which is equivalent to a magnetic field, tends to split the O-2$p$ and Eu-5$d$6$s$ conduction band states. As a consequence, the energy level for unoccupied spin-majority Eu-5$d$6$s$ states, which form the bottom of the CB, is lowered by the exchange splitting $\Delta E_{ex}$. Conversely, the spin-minority level is increased by the same amount. The magnitude of the splitting is described by the exchange interaction $H$ as the following Heisenberg Hamiltonian:[63]

$$H_{ex} = 2\Delta E_{ex} = -2\sum_n J_n(r-R_n)\vec{s}\cdot\vec{S_n}, \quad (6)$$

Where $\vec{s}$ and $\vec{S_n}$ is the electron spin and its neighboring Eu$^{2+}$ ions, respectively, $J_n(r-R_n)$ is the space-dependent exchange constant between the electron and the ion spins. For PM ordering, the summation over Eu$^{2+}$ spins and therefore $\Delta E_{ex}$ will be zero. The unoccupied spin-majority Eu-5$d$6$s$ states then shift to a higher position. As a result, the optical band gap, which is associated with the direct interband transition between the top of the VB and the bottom of the CB, will increase. This explains the experimental observation of the red-shift of the absorption edge from PM to FM ordering.[64] Furthermore, based on the fact that O-2$p$ and Eu-5$d$6$s$ states experience about the same energy shift, we further predict that absorption peaks related to O-$p$ → Eu-5$d$6$s$ transitions will hardly be affected by the FM-PM transitions. We also would like to point out that our calculated exchange splitting of about 0.8 eV is comparable with the value about 0.6 eV determined by spin-resolved x-ray absorption spectroscopy,[7] which is performed on 20 nm thick Eu riched EuO films.

In the above calculations, we assume that the structure of the binary FM EuO is face-centered cubic which possess inversion symmetry, and there is no rumpling between the in-plane anions and cations. To further study the influence of epitaxial strain on the linear optical properties of bulk EuO, we now carry out phonon frequency calculations using the first-principles frozen-phonon method. We change the in-plane lattice constant $a$ within the range 4.873 – 5.495 Å, which is equivalent to applying in-plane strain from -6% to +6%. The out-of-plane lattice constant $c$ is optimized for each $a$. Fig. 3 shows the frequencies of phonon modes versus the in-plane lattice parameter $a$ of EuO. It is clear that when compressive strain is larger than 3.57% ($a < 4.999$ Å) or tensile strain is larger than 4.08% ($a > 5.395$ Å), EuO has imaginary phonon frequencies at the Γ point. Most importantly, the FE state is induced. To be specific, as in the case of rocksalt binary oxides, EuO experiences a splitting of the originally triply degenerated transverse optical (TO) mode into the nondegenerated $A_{2u}$ mode polarized perpendicular to the $xy$ plane and a twofold degenerate $E_u$ mode polarized in the $xy$ plane. With the increase of compressive strain, the $A_{2u}$ mode gradually becomes soft, i.e., its frequency decreases. After about -3.57% strain, the frequency becomes imaginary, indicating the occurrence of an unstable vibration mode as well as PE-FE transition. However, with increase of tensile strain, the $E_u$ mode becomes the soft mode. At about +4.08% strain, a phonon instability appears at the Γ point, suggesting that the ground state of EuO no longer has an inversion center.

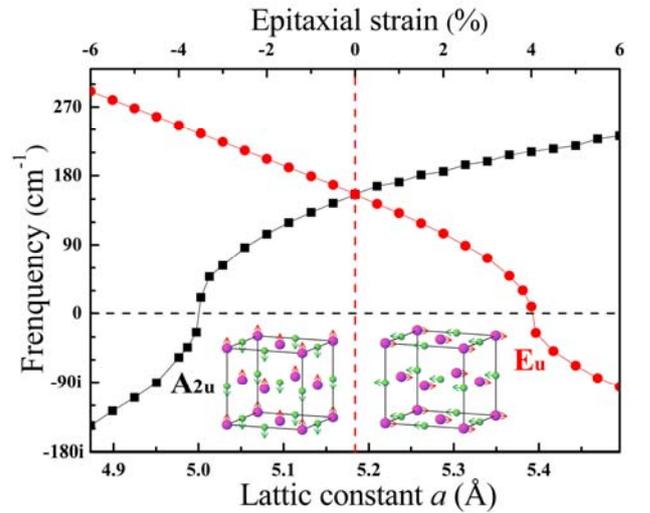

FIG. 3. (Color online) Frequencies of A$_{2u}$ and E$_u$ modes of EuO at the $\Gamma$ point with respect to lattice constant in epitaxial strain. Here the epitaxial strain is defined as $\varepsilon_s = (a-a_0)/a_0 \times 100\%$, where $a_0$ = 5.184 Å is the equilibrium lattice constant, and $a$ is the given in-plane lattice constants.

To search for the stable state of EuO under different strain and confirm the PE-FE phase transition when the phonon instability at the Γ point occurs, we calculate two FE cases, i.e., with 3.6% compressive strain and with 4.1% tensile strain. The final relaxed structure is tetragonal phase ($I4mm$, $C_{4v}$) for 3.6% compressive strain and orthorhombic phase ($Imm2$, $C_{2v}$) for 4.1% tensile strain, compared to cubic phase ($Fm3m$, $O_h$) without strain. Consistent to Kim's work,[41] symmetry lowering of point group and space group can be identified in both cases. The polarization changes from zero at cubic phase to 0.037 C/m$^2$ along the [001] direction with -3.6% epitaxial strain and to 0.050 C/m$^2$ along the [100] (or equivalently [010]) direction with 4.1% in-plane strain, which demonstrates ferroelectricity can be strain engineered in FM EuO.[40]

Epitaxial strain not only leads to structural and PE-FE phase transition, but triggers the change in linear optical response. The imaginary part of dielectric function $\varepsilon_2$, together with the density of states under different strains for EuO, are shown in Fig. 4 to research the influence of epitaxial strain on the linear optical properties of EuO.

For the $Fm3m$ symmetry, there are three imaginary diagonal components: $\varepsilon_{xx}$, $\varepsilon_{yy}$ and $\varepsilon_{zz}$, with $\varepsilon_{xx} = \varepsilon_{yy} = \varepsilon_{zz}$. Closely related to the decrease of point group symmetry, the triply degenerated components split to be doubly degenerated with $\varepsilon_{xx} = \varepsilon_{yy} \neq \varepsilon_{zz}$ under 3.6% compressive strain and all unequal with $\varepsilon_{xx} \neq \varepsilon_{yy} \neq \varepsilon_{zz}$ under 4.1% tensile strain. Here, when the epitaxial strain is applied, the imaginary part of complex dielectric function is regarded as the average of three diagonal parts, which is equivalent to the equation: $\varepsilon_2 = (\varepsilon_{2(xx)} + \varepsilon_{2(yy)} + \varepsilon_{2(zz)}) / 3$.

It is clear that the leading absorption peaks of $\varepsilon_2$ in Fig. 4(a) shift to higher energy levels from +4.1% to -3.6% strain in all spin states, in line with the gradual increase of energy difference between spin-dependent varied orbits from tensile to compressive strain, displayed in Fig. 4(b). In detail, compared with the tiny change of Eu-5d6s spin-majority and spin-minority orbits, the energy of O-2p orbits in both spin states decreases obviously from +4.1% to -3.6% strain, which causes the energy barrier of electron transitions between O-p and Eu-5d6s orbits increases. As the leading peaks E,F,G,H,I are all mainly derived from this kind of interband transition, the blue shift of the leading absorption peaks from tensile to compressive strain under both states can be seen in the optical spectra. The value of $E_g^{opt}$ is 0.62 eV, 0.48 eV, 0.18 eV for the spin-majority channel and 2.98 eV, 2.98 eV, 2.80 eV for the spin-minority channel in the +4.1%, 0% and -3.6% strain states, respectively. From tensile to compressive strain, it is obvious that the red-shift $E_g^{opt}$ in the spin-up case corresponds to the gradual increase of energy difference between highest occupied molecular orbital (HOMO) for spin-majority Eu-4f and lowest unoccupied molecular orbital (LUMO) for spin-majority Eu-5d6s, shown in Fig. 4(b). While, due to the reduced energy of HOMO for O-2p and of LUMO for Eu-5$d$6$s$ from +4.1% to -3.6% strain, the strain-induced differences of $E_g^{opt}$ in the spin-down case are quite small. In all, the energy change of both the leading absorption peaks and the $E_g^{opt}$ in the optical spectra reflects the energy difference between spin-splitting orbits, which means the strain-dependent linear optical responses can be used to explore the strain-dependent density of states, and even strain-dependent band structure.

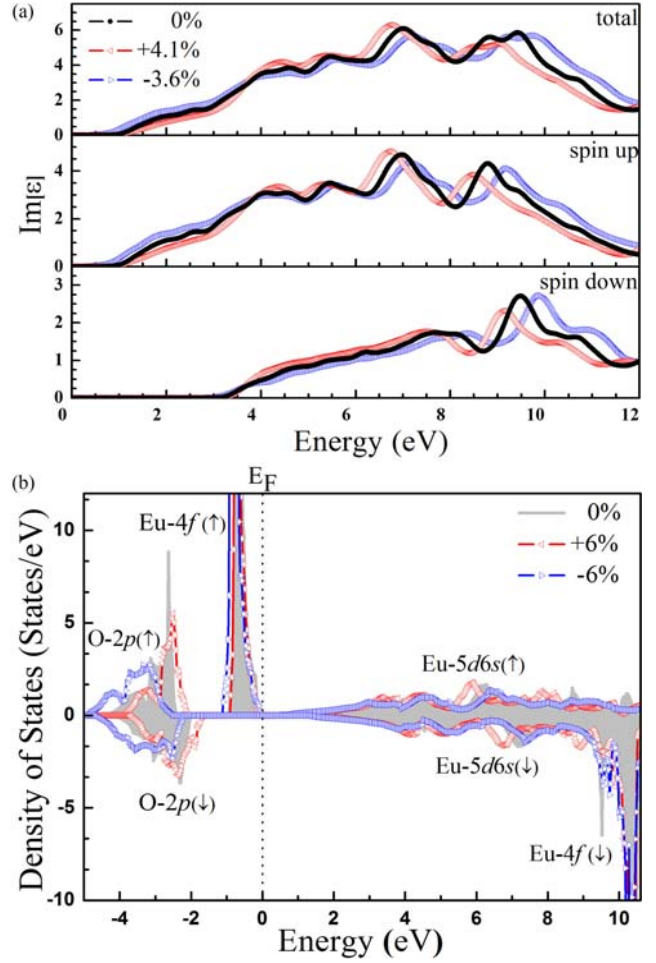

FIG. 4. (Color online) (a) The imaginary part of dielectric constant $\varepsilon$ for EuO with different strain. The solid circle line with black color, left triangle line with red color and right triangle line with blue color represent the case without strain, with 4.1% tensile strain and with 3.6% compressive strain, respectively. (b) The corresponding density of states with different strain. The zero of energy is at the Fermi level, $E_F$.

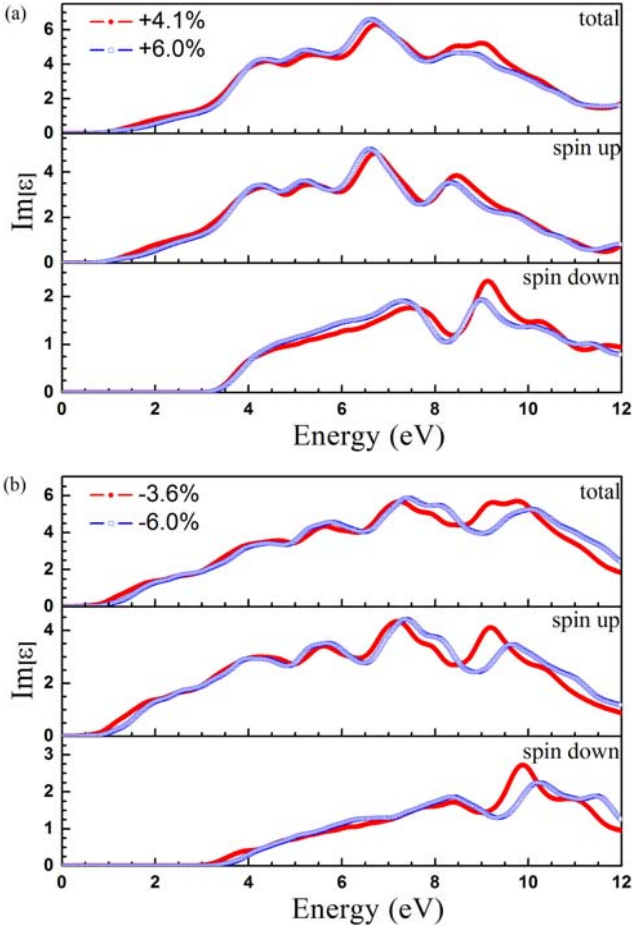

FIG. 5. (Color online) The strain effect of the imaginary part of dielectric constant $\varepsilon$ for EuO (a) under tensile strain and (b) under compressive strain.

As the linear optical response is sensitive in different strain states, what will happen when the strain value changes? In order to explore the strain effect, we calculate the imaginary part of dielectric constant with 6.0% compressive strain and 6.0% tensile strain, and compare them to -3.6% and +4.1% cases, respectively. It can be clearly seen in Fig. 5 that under tensile strain, the red shift of the leading absorption peaks and $E_g^{opt}$ will occur with the increase of strain value. Conversely, when compressive strain is applied, the energy of the leading absorption peaks and $E_g^{opt}$ will increase. To summarize, relative to the case without strain, the larger the epitaxial strain applied is, the bigger the deviation of the absorption peaks will be.

The more interesting thing is, as the ferroelectricity is induced and the space inversion symmetry is broken due to epitaxial strain, EuO film will present the nonlinear optical properties. As examples, we calculate the SHG coefficients of EuO under 3.6% compressive and 4.1% tensile strain, and compare them to the cases under +6.0% and -6.0% strain. Again, the results are analyzed spin dependently, as shown below.

Since the point group is $C_{2v}$ when tensile strain applied, there are five independent SHG components: $\chi^{(2)}_{xxx}$, $\chi^{(2)}_{xyy}$, $\chi^{(2)}_{xzz}$, $\chi^{(2)}_{yyx}$, $\chi^{(2)}_{zzx}$. Whereas due to the point group of $C_{4v}$ with compressive strain applied, there are five nonvanishing SHG components, i.e. $\chi^{(2)}_{xxz}$, $\chi^{(2)}_{yyz}$, $\chi^{(2)}_{zxx}$, $\chi^{(2)}_{zyy}$ and $\chi^{(2)}_{zzz}$. The twofold symmetry $z$-axis gives rise to the equivalent components, written as $\chi^{(2)}_{xxz} = \chi^{(2)}_{yyz}$ and $\chi^{(2)}_{zxx} = \chi^{(2)}_{zyy}$. Furthermore, according to the so-called Kleinman symmetry[65] which demands $\chi^{(2)}_{xyy} = \chi^{(2)}_{yyx}$, $\chi^{(2)}_{xzz} = \chi^{(2)}_{zzx}$ and $\chi^{(2)}_{xxz} = \chi^{(2)}_{zxx}$ in the static limit, the static value of SHG susceptibility (in unit of $10^{-9}$ esu) can be described as $\chi^{(2)}_{xxx}$ = 11.88, $\chi^{(2)}_{xyy}$ = -2.67, $\chi^{(2)}_{xzz}$ = -0.87 under +4.1% strain, and $\chi^{(2)}_{xxz}$ = -13.22, $\chi^{(2)}_{zzz}$ = -36.57 under -3.6% strain, respectively. When the value of the epitaxial strain increases, the SHG susceptibility in the static limit will change as $\chi^{(2)}_{xxx}$ = 169.76, $\chi^{(2)}_{xyy}$ = -13.19, $\chi^{(2)}_{xzz}$ = -4.36 under 6.0% tensile strain, and $\chi^{(2)}_{xxz}$ = -45.03, $\chi^{(2)}_{zzz}$ = 17.60 under 6.0% compressive strain. It's interesting to find that with the enhancement of the epitaxial strain, the absolute value of SHG susceptibility will increase, except the $\chi^{(2)}_{zzz}$ component. Thanks to our spin-dependent research, the cause for sign reversal of it can be obtained. The contributions for the $\chi^{(2)}_{zzz}$ component in the static limit from different spin states are of opposite sign. The bigger the value of the compressive strain is, the larger the absolute value of the positive spin-down channel will be. Meanwhile, that of the negative spin-up state will decrease. With a relatively small strain, the contribution from the negative spin-up state is dominant. When the 6.0% compressive strain is applied, the absolute value of the spin-down case is larger than that of the spin-up case, which triggers sign reversal of the $\chi^{(2)}_{zzz}$ component.

In order to research the frequency dependent curves of SHG coefficients and explore the strain effect, we plot the real and imaginary part of component $\chi^{(2)}_{xxx}$ with 4.1% and 6.0% tensile strain, and that of $\chi^{(2)}_{zzz}$ with 3.6% and 6.0% compressive strain in Fig. 6. Note that, due to the polarization direction, the electric-dipole transitions along $x$-axis under tensile strain and along $z$-axis under compressive strain can only occur between energy levels with identical symmetry, which makes the components $\chi^{(2)}_{xxx}$ and $\chi^{(2)}_{zzz}$ exceptional. For all the curves within the low energy range, it is obvious that, compared with spin-up channels, the curves of spin-down channels are relatively flat and without sharp peaks. This indicates that in the infrared and visible light region, the SHG coefficients are mainly determined by the virtual-electron and virtual-hole process of spin-up states. It is caused by the fact that the energy differences between VB and CB for spin-minority states are considerably larger than those of spin-majority states, which plays a more significant role in the nonlinear optical response. The comparison between different strain values shows that with the increase of epitaxial strain, the nonlinear optical response will be more significant, partially due to farer from the inversion symmetry. As the SHG susceptibility is sensitive to lattice structure, the strain effect is quite remarkable compared with the linear optics, especially in ultraviolet region. The phenomenon described above can also be found in other nonvanishing components that are not shown here.

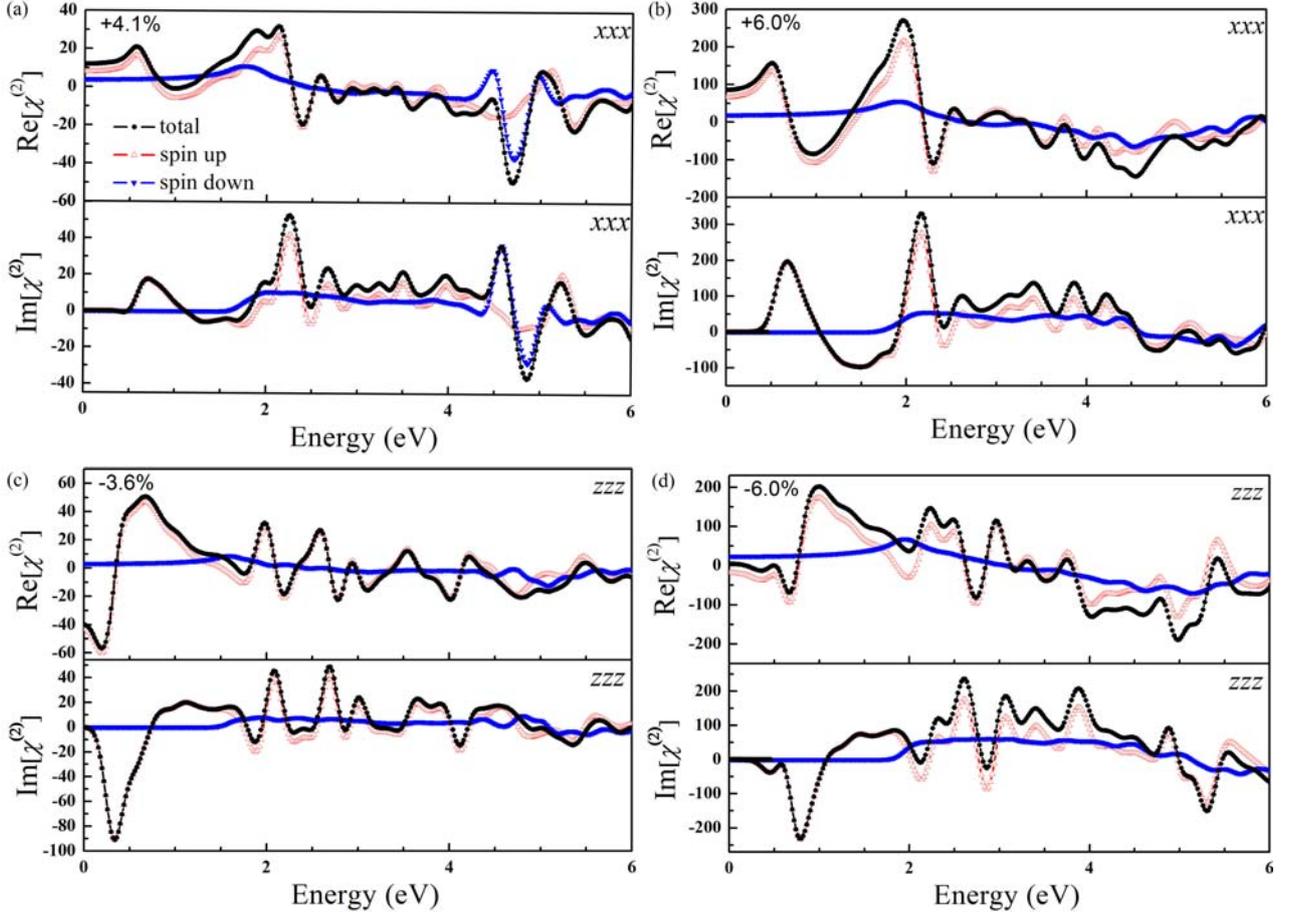

FIG. 6. (Color online) The real and imaginary part of SHG susceptibilities (a) $\chi^{(2)}_{xxx}$ with 4.1% tensile strain, (b) $\chi^{(2)}_{xxx}$ with 6.0% tensile strain, (c) $\chi^{(2)}_{zzz}$ with 3.6% compressive strain and (d) $\chi^{(2)}_{zzz}$ with 6.0% compressive strain in units of $10^{-9}$ esu. The solid square line with black color, up triangle line with red color and down triangle line with blue color are for total, spin-up and spin-down states, respectively.

## IV. CONCLUSION

In conclusion, we have studied the spin-dependent linear and nonlinear optical properties of ferromagnetic EuO by using DFT calculations. Spin-resolved spectrum is obtained and the origin of its absorption peaks is analyzed, which is necessary to research electron transition in different spin states. In addition, we find that exchange interaction causes the spin-splitting of Eu-5$d$6$s$ conduction band, as well as the red-shift of the absorption edge from paramagnetic to ferromagnetic ordering. Moreover, we have studied the influence of epitaxial strain on the optical properties of EuO. With the application of different epitaxial strains, the paraelectric-ferroelectric transition can be induced, i.e. EuO could be multiferroic. From tensile to compressive strain, the leading absorption peaks in the optical spectra tend to have blue-shift, whereas the optical band gap in spin-up state will decrease, in line with the energy difference between spin-dependent varied obits. With the increase of strain value, the energy shift of the leading absorption peaks will be more significant. The analyses on the spin-dependent nonlinear optical properties indicate that SHG susceptibilities of multiferroic EuO in the infrared and visible light region are mainly from the contribution of spin-majority states. The strain effect is obvious in the static limit, as well as the dispersion curves.

# ACKNOWLEGEMENTS

This work was supported by the 973 Program No. 2013CB922301, 2014CB921104, 2011CB922101, the NSF of China (Grant Nos. 61125403, 91122035, 11174124), Program of Shanghai Subject Chief Scientist, and Fundamental Research Funds for the central universities (ECNU). Computations were performed at the ECNU computing center.